\documentclass{article}
\usepackage{spconf,amsmath,graphicx}
\usepackage{cite}
\usepackage{setspace}
\usepackage{booktabs}
\usepackage{graphicx}

\makeatother
\begin{document}
\title{LightVessel: Exploring Lightweight Coronary Artery Vessel Segmentation via Similarity Knowledge Distillation}
%
\name{Hao Dang$^{1,\ast}$\thanks {$\ast$ These authors contributed equally to this work.}, Yuekai Zhang$^{2, \ast}$, Xingqun Qi$^{2, \dagger}$, Wanting Zhou$^{2}$, Muyi Sun$^{3}$\thanks {$\dagger$ Corresponding author.}
\thanks{This paper is funded by China Postdoctoral Science Foundation (2022M713362) and Henan Province Key Specialized Research and Development Breakthrough Program (No.222102210028).}
\address{$^{1}$School of Information Technology, Henan University of Chinese Medicine, Zhengzhou, China \\
$^{2}$Beijing University of Posts and Telecommunications, Beijing, China \\
$^{3}$Institute of Automation, Chinese Academy of Sciences, Beijing, China}
}

%
\maketitle
\begin{abstract}
\setlength{\baselineskip}{11.2pt}
In recent years, deep convolution neural networks (DCNNs) have achieved great prospects in coronary artery vessel segmentation.
However, it is difficult to deploy complicated models in clinical scenarios since high-performance approaches have excessive parameters and high computation costs. 
To tackle this problem, we propose \textbf{LightVessel}, a Similarity Knowledge Distillation Framework, for lightweight coronary artery vessel segmentation. 
Primarily, we propose a Feature-wise Similarity Distillation (FSD) module for semantic-shift modeling. Specifically, we calculate the feature similarity between the symmetric layers from the encoder and decoder. Then the similarity is transferred as knowledge from a cumbersome teacher network to a non-trained lightweight student network.
Meanwhile, for encouraging the student model to learn more pixel-wise semantic information, we introduce the Adversarial Similarity Distillation (ASD) module.
Concretely, the ASD module aims to construct the spatial adversarial correlation between the annotation and prediction from the teacher and student models, respectively. 
Through the ASD module, the student model obtains fined-grained subtle edge segmented results of the coronary artery vessel.
Extensive experiments conducted on Clinical Coronary Artery Vessel Dataset demonstrate that LightVessel outperforms various knowledge distillation counterparts.

\end{abstract}
\begin{keywords}
Coronary angiography, segmentation, knowledge distillation, lightvessel
\end{keywords}
\section{Introduction}
\label{sec:intro}
\setlength{\baselineskip}{11.2pt}

\begin{figure}[t]
\centering
\includegraphics[width=8cm]{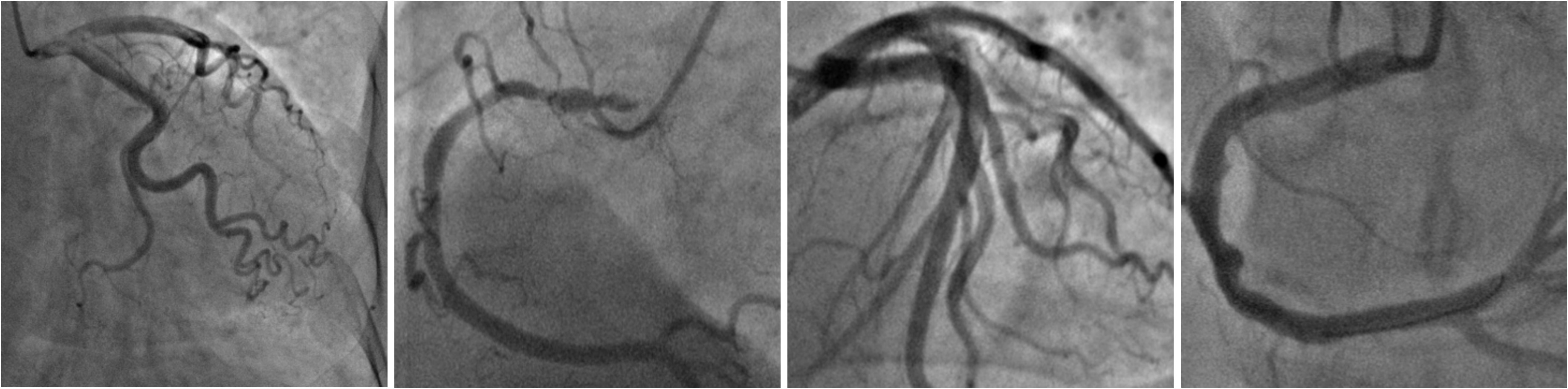}
\caption{The samples from X-ary coronary angiography images with low contrast, fuzzy vessel boundary, complex background, and structural interference.}
\label{fig:figure1}
\vspace{-0.5cm}
\end{figure}

Coronary artery disease (CAD) is one leading cause of mortality and its prevalence is projected to increase worldwide gradually \cite{ref1,ref2}. 
To obtain accurate diagnosis results, coronary artery vessel segmentation is imperative from X-ary angiography images shown in Figure \ref{fig:figure1}.  
There are various endeavors\cite{ref3,ref4,ref5,ref6,ref7}  which engage in vessel segmentation. 
CNN has become a milestone for this task in recent years. 
However, these methods always embrace complex computation and massive parameters, which are degraded or even incompetent in the scenes with limited computational resources. 
Meanwhile, the size of angiography images is usually larger than natural images. Therefore, the trade-off between performance and cost requires to be considered.

To tackle the above dilemma, lightweight networks are proposed to alleviate the burden.
Model compression is one of the main directions to construct lightweight model, which is roughly categorized into three classes: pruning \cite{ref8}, quantization \cite{ref9}, and knowledge distillation \cite{ref10, ref11, ref12,ref13, ref14,ref15}. 
Pruning and quantization try to adjust and modify the architecture of the networks, which leads to limitations on robustness and generalization. 
Knowledge distillation explores the knowledge transfer from a  high-performance network to a non-trained lightweight network. 
This approach realizes high efficiency and accuracy, which also avoids designing new structures. 
Therefore, in this paper, we explore knowledge distillation to circumvent the trade-off between accuracy and efficiency in coronary artery vessel segmentation.

\graphicspath{{Images/}}

\begin{figure*}[t]
\centering
\includegraphics[width=0.75\textwidth]{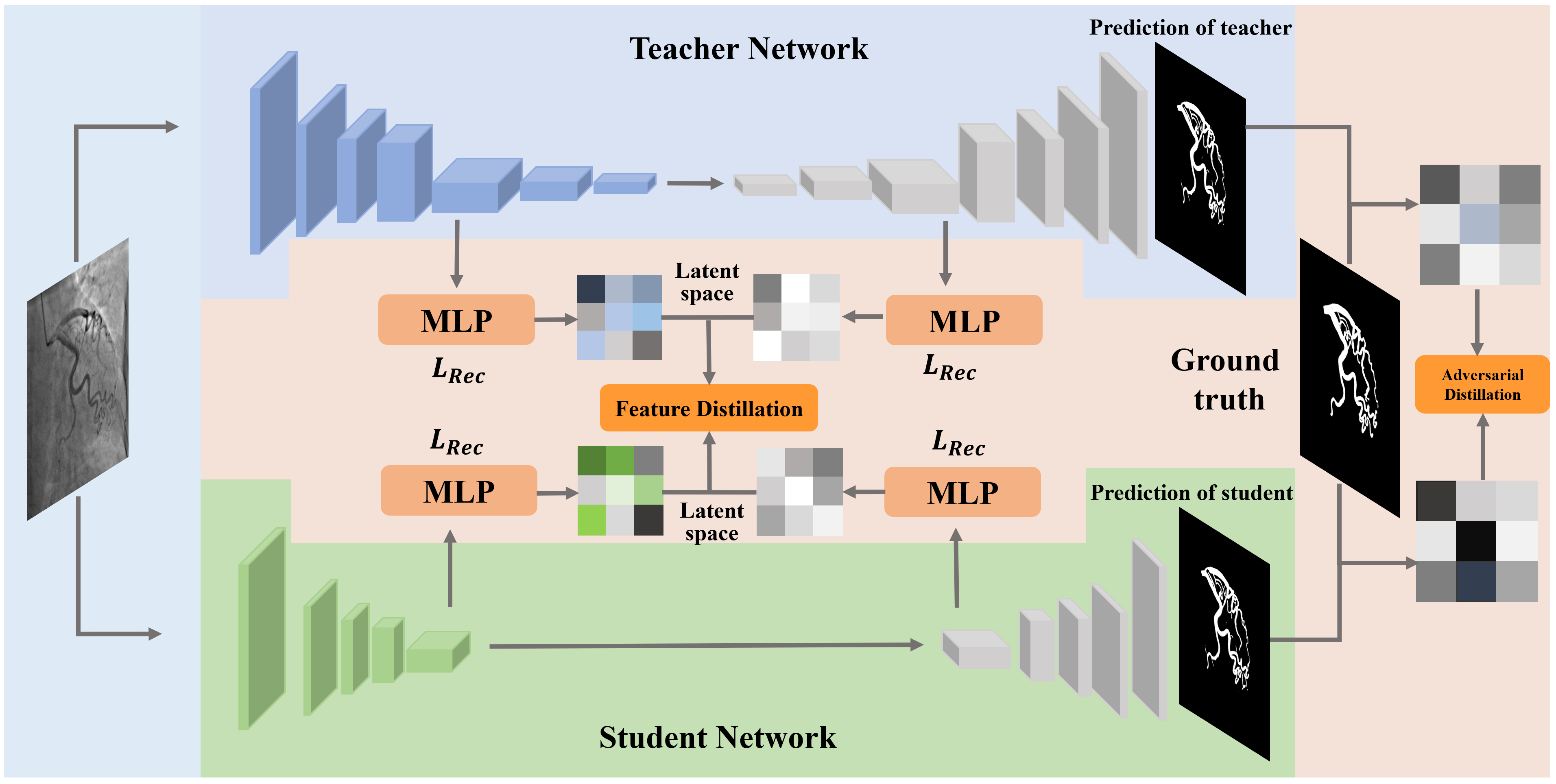}
\caption{The pipeline of our proposed LightVessel. The feature distillation is the FSD module, which is committed to transmitting knowledge from a cumbersome teacher network to a non-trained lightweight student network. The adversarial distillation aims to compute the similarity between the predicted results and annotations.}
\label{fig:figure2}
\vspace{-0.5cm}
\end{figure*}

Motivated by the above observation, we propose \textbf{LightVessel} for coronary artery segmentation. The framework of LightVessel is shown in Figure \ref{fig:figure2}, which includes two modules: Feature-wise Similarity Distillation (FSD) and Adversarial Similarity Distillation(ASD).  In the FSD module, we calculate feature similarity and distill it from the tanglesome teacher network to lightweight student network, which enhances the semantic correlation extraction ability of the student network. In the ASD module, we calculate the adversarial similarity through the predicted results and annotations. The semantic representations from the teacher and student are both associated with the semantic annotations, which promote the student network to obtain fine-grained semantic features.

The main contributions are as follows: 
(1) We propose LightVessel, a Similarity Distillation Framework, to perform  efficient coronary artery vessel segmentation. 
(2) We design two collaborative modules, Feature-wise Similarity Distillation (FSD) and Adversarial Similarity Distillation(ASD), which perform high-quality and fine-grained semantic feature distillation.
(3) The ablation and generalization experiments are implemented on Clinical Coronary Artery Vessel Dataset to verify the effectiveness and robustness of LightVessel. 
\vspace{-0.3cm}

\section{Related works}
\label{sec:format}
\setlength{\baselineskip}{11.2pt}

\subsection{Coronary Artery Vessel Segmentation}
\label{ssec:subhead}

\graphicspath{{Images/}}

With the dramatic development of CNN, coronary artery vessel segmentation has made great progress in recent years. Samuel \emph{et al.}\cite{ref13} leveraged a VSSC network to segment vessels from coronary angiographic images. 
Zhu \emph{et al.}\cite{ref4} used PSPNet to realize a parallel multi-scale CNN solving the problems of the complex structure of angiography images. Jun \emph{et al.}\cite{ref14} proposed an encoder-decoder architecture called T-Net, which consists of several small encoders. T-Net overcomes the limitation of U-Net\cite{ref15}, which has only one set of connection layers between encoder and decoder blocks. 
However, the prominent limitation of these models is the high complexity, which is caused by the excessive parameters and computational costs. 
Inspired by the above, we propose a novel lightweight method based on knowledge distillation.

\subsection{Knowledge Distillation}
\label{ssec:subhead}

\vspace{-0.1cm}
Knowledge distillation is one of the most popular lightweight algorithms for CNN. 
Recently, this method has been applied to medical image analysis. 
Zhou \emph{et al.} \cite{ref16} proposed a knowledge distillation method to leverage both labeled and unlabeled data for overlapping cervical cell instance segmentation. 
Dou \emph{et al.} \cite{ref17} designed a learning scheme with a novel loss inspired by knowledge distillation, to implement multi-modal medical image segmentation. 
Hu \emph{et al.} \cite{ref18} established a knowledge distillation network to transfer knowledge from a trained multi-modal network to a mono-modal, which achieves accurate brain tumor segmentation. 
However, the challenges of these methods are two-fold. 
Firstly, the previous models ignore the whole transformation process of the teacher network for semantic information. 
Then, there is mismatching of feature dimensions between complex teacher models and lightweight student models. 
In this paper, we establish LightVessel method to circumvent these limitations.

\vspace{-0.3cm}

\section{Methodology}
\label{sec:pagestyle}
\setlength{\baselineskip}{11.2pt}

\subsection{Overview}
\label{ssec:subhead}

As illustrated in Figure \ref{fig:figure2}, the LightVessel includes Feature-wise Similarity Distillation modules (FSD) and Adversarial Similarity Distillation modules (ASD).
Specifically, the coronary angiography images $X$ are input to the teacher network $T$ and the student network $S$, simultaneously.
To construct the semantic-shift of corresponding feature maps, the FSD module leverages the MLP to map the feature of the encoder and decoder into latent space. Then the feature similarity is calculated and distilled to $S$. 
Meanwhile, to establish pixel-level spatial correlations, the ASD module constructs adversarial similarity between the predicted results and annotations. The predicted results are projected into semantic space to ensure spatial consistency of the feature similarity with the ASD module.

\subsection{Feature-wise Similarity Distillation}
\label{ssec:subhead}
\begin{figure}[t]
\centering
\includegraphics[width=0.4\textwidth]{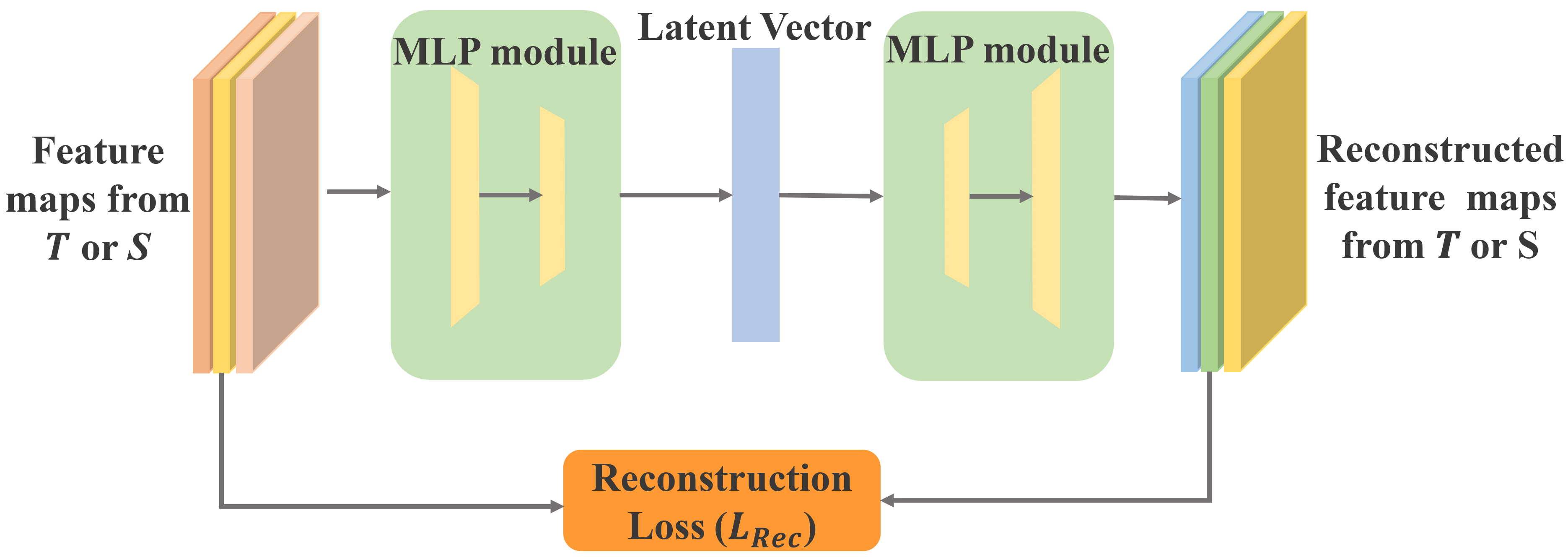}
\caption{The structure of MLP module.}
\label{fig:figure3}
\vspace{-4mm}
\end{figure}

The FSD module is designed to transmit semantic-ware feature information ($knowledge$) from the cumbersome $T$ to the lightweight $S$. 
Concretely, the corresponding feature maps from the encoder and decoder are projected into the latent space through MLP depicted in Figure \ref{fig:figure3}. The MLP aims to obtain the dimension-consistency latent vectors from the feature maps of teacher and student models, respectively. In the training phase, we first obtain the scratch-trained teacher and student models, then we utilize the $\mathcal{L}_{Rec}$ for training MLP. The reconstruct loss $\mathcal{L}_{Rec}$ is defined as:

\graphicspath{{Images/}}

\begin{figure}[t]
\centering
\includegraphics[width=0.43\textwidth]{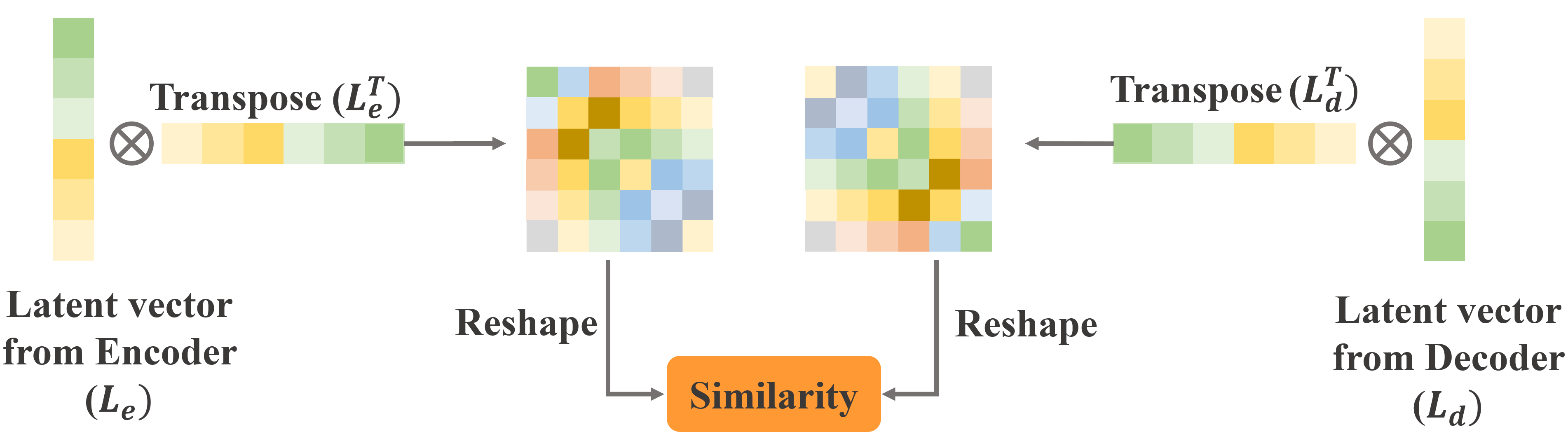}
\caption{Feature-wise Similarity Distillation module.}
\label{fig:figure4}
\vspace{-0.5cm}
\end{figure}

\begin{equation}
\mathcal{L}_{Rec}=\big|\big| Fea_{in}-Fea_{out} \big|\big|_1
\end{equation}
Where $Fea_{in}$ denotes the features from $T$ and $S$, and $Fea_{out}$ illustrates the reconstructed features with MLP.
Further, the similarity distillation is illustrated in Figure \ref{fig:figure4}. Concretely, the symmetric latent vectors $L_{e/d}=[l_1, l_2,\ldots,l_c]\in R^{1\times512}$ are transposed ($L_{e/d}^T$) and multiply original $L_{e/d}$. The output is called $L_{tra}\in R^{512\times512}$. Meanwhile,  We reshape $L_{tra}$ ($512\times512$ vector) into a 1-dimensional vector ($L_{1-D}$). Next, we calculate cosine similarity to obtain the feature dimension similarity maps. The cosine similarity is defined as: 

\graphicspath{{Images/}}

\begin{equation}
Cos(L_e, L_d) =\frac{L_e\cdot L_d}{\big|\big| L_e \big|\big|_2 \cdot \big|\big| L_d \big|\big|_2}
\end{equation}
Where $L_e$ and $L_d$ denote the latent vectors from the encoder and decoder, respectively. The cosine similarity is calculated through two latent vectors. Thus, we obtain the semantic changes from the encoder to the decoder in $T$ and $S$. Given the cosine similarity maps of the corresponding feature dimension from $T$ and $S$, the $L2$ loss function is leveraged to constrain the cosine similarity maps. It is defined as:

\begin{equation}
\mathcal{L}_{FSD}=\big|\big| Cos_{tea} - Cos_{stu} \big|\big|_2
\end{equation}

\begin{table}[t]
\centering
\caption{The comparisons with SOTA KD methods. $\downarrow$ indicates the lower is better, and $\uparrow$ indicates the higher is better.}
\resizebox{\linewidth}{!}{
\renewcommand{\arraystretch}{1.1} 
\begin{tabular}{c c c c c c c c} 
\toprule
\textbf{Models}           & \textbf{ACC}$\uparrow$    & \textbf{Se}$\uparrow$     & \textbf{AUC}$\uparrow$     & \textbf{mIOU}$\uparrow$   & \textbf{F1-score}$\uparrow$ & \textbf{FLOPs}$\downarrow$  & \textbf{Parms}$\downarrow$   \\ \hline
\textbf{Teacher}          & \textbf{0.9835} & \textbf{0.8671} & \textbf{0.9929} & \textbf{0.8112} & \textbf{0.7801}   & \textbf{78.76G} & \textbf{26.489M} \\
\textbf{Student-scratch}  & \textbf{0.9804} & \textbf{0.8155} & \textbf{0.9865} & \textbf{0.7822} & \textbf{0.7378}   & \textbf{0.804G} & \textbf{4.682M}  \\
SoftKD\cite{ref12}                    & 0.9820          & 0.8548          & 0.9919          & 0.7985          & 0.7620            & 0.804G          & 4.682M           \\
ATKD\cite{ref19}                      & 0.9812          & 0.8402          & 0.9904          & 0.7909          & 0.7509            & 0.804G          & 4.682M           \\
IFVD\cite{ref20}                      & 0.9830          & 0.8502          & 0.9913          & 0.8056          & 0.7721            & 0.804G          & 4.682M           \\
SKD\cite{ref21}                       & 0.9822          & 0.8544          & 0.9918          & 0.7800          & 0.7641            & 0.804G          & 4.682M           \\ \hline
\textbf{Student-Ours} & \textbf{0.9835} & \textbf{0.8620} & \textbf{0.9936} & \textbf{0.8106} & \textbf{0.7791}   & \textbf{0.804G} & \textbf{4.682M} \\
\bottomrule
\end{tabular} 
\label{tab:table1}
}
\end{table}

\graphicspath{{Images/}}
\begin{figure}[t]
\centering
\includegraphics[width=0.48\textwidth]{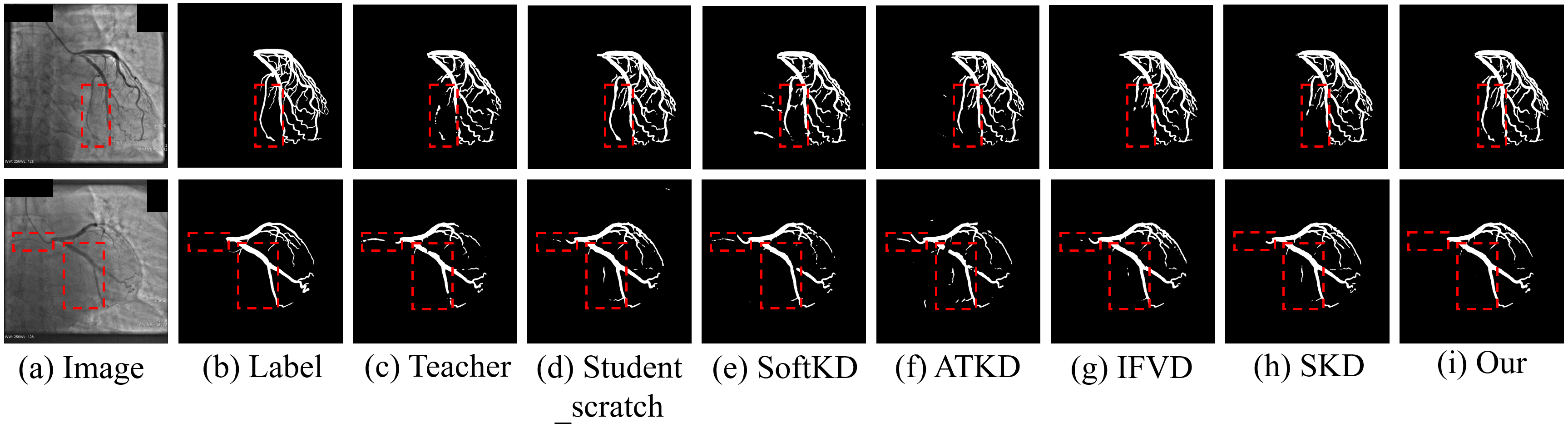}
\caption{The results compared with SOTA KD methods. Please \textbf{zoom in for better visualization}.}
\label{fig:figure5}
\vspace{-0.5cm}
\end{figure}

\subsection{Adversarial Similarity Distillation}
\label{ssec:subhead}

To further enhance $S$ to learn more refined semantic information from $T$, we propose the ASD module. More specifically, the ASD module calculates the pixel similarity between the prediction results of teacher network and annotations. 
Thus, the ASD module supervises the spatial correlation between predictions and annotations for pixel-level feature realignment. To retain abundant spatial semantic, the Euclidean distance is implemented as a metric. And it is defined as:
\begin{equation}
Euc(T,G)=\sqrt{\sum_{i=1}^n (T_i -G_i)^2}
\end{equation}
Where $T$ denotes the predictions of teacher, $G$ shows the annotations. 
The $L2$ loss function is carried out to supervise the pixel-wise knowledge transfer from $T$ to $S$.
It is defined as:

\begin{equation}
\mathcal{L}_{ASD} = \big|\big|Euc_{tea} -Euc_{stu}  \big|\big|_2
\end{equation}

In addition, we utilize the labels and predictions of student to define the cross-entropy loss function. It is defined as:
\begin{equation}
\mathcal{L}_{CE}=-\frac{1}{N} \sum(y_ilog(\overline{y_i})+(1-y_i)log(1-\overline{y_i}))
\end{equation}
Where $y_i$ denotes ground truth, $\overline{y_i}$ shows predicted result. $N$ describes the number of samples.
Last, the $\mathcal{L}_{CE}$ is combined with $\mathcal{L}_{FSD}$ and $\mathcal{L}_{ASD}$ to construct the overall objective $\mathcal{L}_{total}$:
\begin{equation}
\mathcal{L}_{total}=\mathcal{L}_{CE}+\mathcal{L}_{FSD}+\mathcal{L}_{ASD}
\end{equation}

\section{Experiments and results}
\label{sec:typestyle}
\setlength{\baselineskip}{11.2pt}
\subsection{Experiments Setup}
\label{ssec:subhead}

\textbf{Dataset.} We employ Clinical Coronary Artery VesseDataset for experiments. 
The original dataset includes 240 images (992×992) and is augmented into 3864 images (256×256) for the experiment. The training and testing sets include 3200 and 640 images, respectively. In our experiments, we adopt patch-based method expressed as the original image is cropped into 256$\times$256 patches.\\
\textbf{Implemention Details and Evaluation Metrics.}
In the training phase, the initial learning rate is set as $1e-3$.
The total epoch is set to 200 with the batch size of 16. All the models are implemented on the Pytorch \cite{ref22} platform with 2 Nvidia Titan XP GPUs. To verify the segmentation performance, we adopt five metrics: accuracy, sensitivity, AUC, mIOU, and F1-score. Furthermore, we leverage the sum of point operations (FlOPs) and the number of network parameters (Params) to measure lightweight.

\subsection{Experiment Results}

\subsubsection{Contrastive Analysis with KD Methods}
\label{ssec:subhead}

In this section, we compare LightVessel with four state-of-the-art knowledge distillation approaches including SoftKD\cite{ref12}, ATKD\cite{ref19}, IFVKD\cite{ref20}, and SKD\cite{ref21}.
The four methods are performed with consistent original parameters.
The backbone network is U-net. Specifically, the encoder is stacked with attention-based SKblock to establish a teacher network with large parameters. And the encoder of U-net is stacked with the inverted bottleneck block of lightweight MobileNetV2\cite{ref23} to construct the student network with slight parameters. The detailed experiment results are depicted in Table \ref{tab:table1}. 

As reported in Table \ref{tab:table1}, our proposed LightVessel outperforms various knowledge distillation counterparts. Especially for the sensitivity, the student network increased by nearly 0.05. The other metrics have also been greatly improved, and even AUC has exceeded the teacher network. It is worth noting that our method achieves performance improvement without increasing the computation costs and parameters compared with the original student network. We also show the qualitative segmentation effectiveness described in Figure \ref{fig:figure5}. The performance of student network(i) with LightVessel has significantly improved semantic information extraction ability. Meanwhile, compared with the other four methods, the LightVessel also has better performance.

\vspace{-0.3cm}
\subsubsection{Feature-wise Similarity Distillation}
\label{ssec:subhead}

We report the effectiveness of the FSD module in Table \ref{tab:table2}. The performance of the student network with FSD has a large improvement. Especially for the sensitivity, our student network with FSD increased by 0.034. Figure 6 illustrates the ablation experiment visualization result of the student network with FSD module (e). The segmentation ability of the student network with FSD module has been significantly optimized.

\vspace{-0.3cm}
\subsubsection{Adversarial Similarity Distillation}
\label{ssec:subhead}

\begin{table}[t]
\centering
\caption{The ablation experiments of FSD and ASD modules}
\resizebox{.45\textwidth}{!}{
\renewcommand{\arraystretch}{1.1} 
\begin{tabular}{c c c c c c c c} \toprule
\textbf{Models}          & \textbf{ACC}$\uparrow$      & \textbf{Se}$\uparrow$      & \textbf{AUC}$\uparrow$     & \textbf{mIOU}$\uparrow$    & \textbf{F1-score}$\uparrow$      \\ \hline
Teacher                 & 0.9835          & 0.8671          & 0.9929          & 0.8112          & 0.7801          \\
Student-scratch         & 0.9804          & 0.8155          & 0.9865          & 0.7822          & 0.7378          \\
\textbf{S w/ FSD}       & \textbf{0.9826} & \textbf{0.8495} & \textbf{0.9917} & \textbf{0.8021} & \textbf{0.7670}  \\
\textbf{S w/ FSD + ASD} & \textbf{0.9835} & \textbf{0.8620} & \textbf{0.9936} & \textbf{0.8106} & \textbf{0.7791} \\  \bottomrule
\end{tabular} 
\label{tab:table2}
}
\end{table}

\graphicspath{{Images/}}

\begin{figure}[t]
\centering
\includegraphics[width=0.4\textwidth]{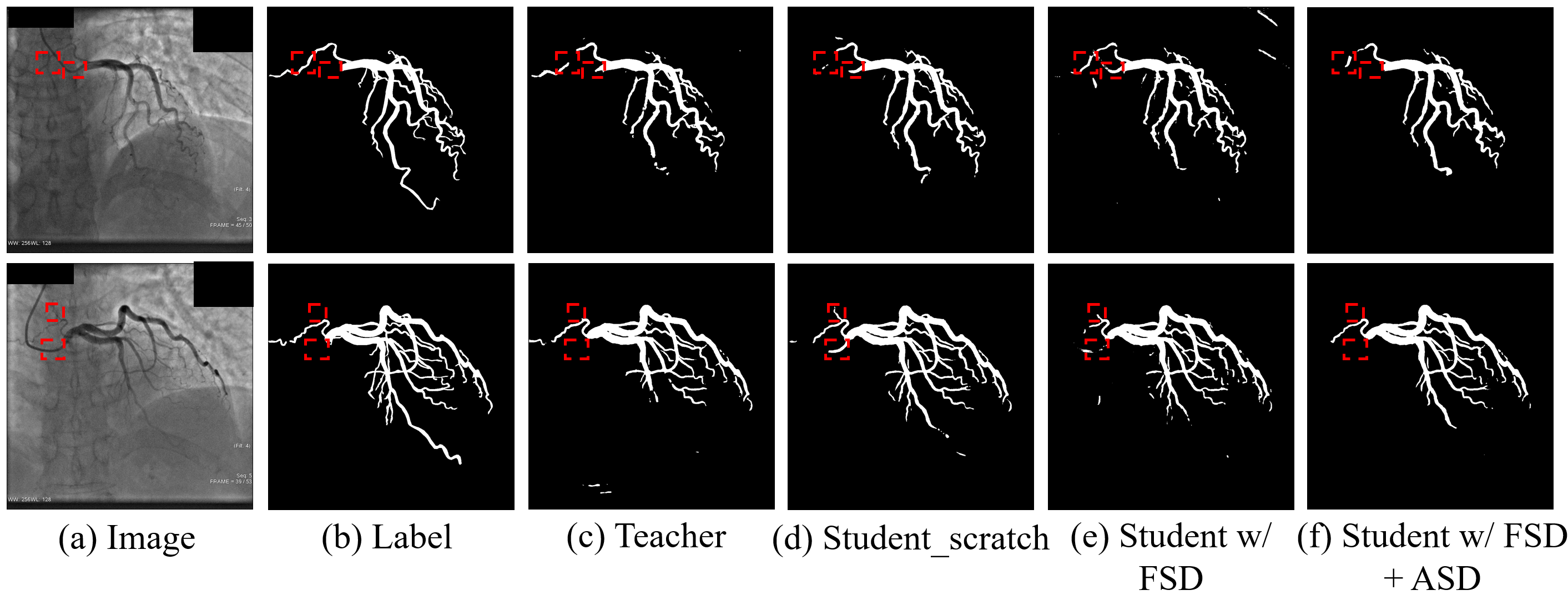}
\vspace{-0.3cm}
\caption{The ablation analysis of FSD and ASD modules. Please \textbf{zoom in for better visualization}.}
\label{fig:figure6}
\vspace{-0.5cm}
\end{figure}

As shown in Table \ref{tab:table2}, the quantitative analysis is carried out to validate the rationality and effectiveness of ASD module. Especially for the sensitivity, the student network with FSD and ASD modules (f) increased by nearly 0.05.  Figure 6 also showcases the detailed qualitative results. The subtle edges of the vessels have been significantly optimized.

\begin{table}[t]
\centering
\caption{Generalization verification. St- (MobileV2, ENet, ERFNet) -scratch denotes three student networks without KD.}
\resizebox{\linewidth}{!}{
\renewcommand{\arraystretch}{1.1} 
\begin{tabular}{c c c c c c c c} \toprule
\textbf{Models}                  & \textbf{ACC}$\uparrow$     & \textbf{Se}$\uparrow$    & \textbf{AUC}$\uparrow$    & \textbf{mIOU}$\uparrow$   & \textbf{F1-score}$\uparrow$ & \textbf{FLOPs}$\downarrow$  & \textbf{Params}$\downarrow$  \\ \hline
\textbf{Teacher}                 & \textbf{0.9835} & \textbf{0.8671} & \textbf{0.9929} & \textbf{0.8112} & \textbf{0.7801}   & \textbf{78.76G} & \textbf{26.489M} \\
St-Mobilev2-scratch           & 0.9804          & 0.8155          & 0.9865          & 0.7822          & 0.7378            & 0.804G          & 4.682M           \\
\textbf{St-Mobilev2-our} & \textbf{0.9835} & \textbf{0.8620} & \textbf{0.9936} & \textbf{0.8106} & \textbf{0.7791}   & \textbf{0.804G} & \textbf{4.682M}  \\ \hline
St-ENet-scratch                  & 0.9816          & 0.8117          & 0.9892          & 0.7895          & 0.7484            & 0.516G          & 0.349M           \\
\textbf{St-ENet-our}        & \textbf{0.9831} & \textbf{0.8668} & \textbf{0.9934} & \textbf{0.8082} & \textbf{0.7758}   & \textbf{0.516G} & \textbf{0.349M}  \\ \hline
St-ERFNet-scratch                & 0.9812          & 0.8152          & 0.9900          & 0.7876          & 0.7458            & 3.22G           & 2.063M           \\
\textbf{St-ERFNet-our}      & \textbf{0.9833} & \textbf{0.8696} & \textbf{0.9933} & \textbf{0.8075} & \textbf{0.7750}   & \textbf{3.22G}  & \textbf{2.063M}  \\ \bottomrule
\end{tabular} 
\label{tab:table3}
}
\end{table}

\subsubsection{Generalization Analysis}
\label{ssec:subhead}
\begin{figure}[h]
\centering
\includegraphics[width=0.48\textwidth]{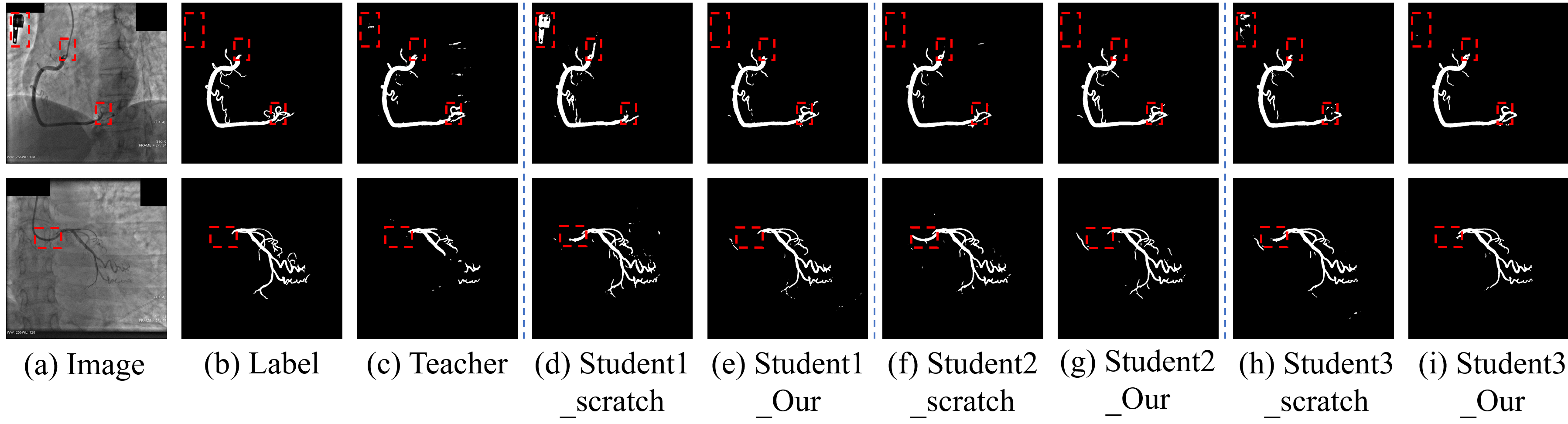}
\vspace{-0.5cm}
\caption{The visualization results of generalization experiments. Student1, student2, and student3 are three different networks based on MobileV2, ENet, and ERFNet. Please \textbf{zoom in for better visualization}.}
\label{fig:figure7}
\vspace{0cm}
\end{figure}

To verify the generalization of our proposed methods, we construct different student models based on ENet\cite{ref24} and ERFNet\cite{ref25}. The experiment results are depicted in Table \ref{tab:table3}. Figure \ref{fig:figure7} shows the visualization results of generalization experiments. 
LightVessel has achieved significant performance compared with other student networks. This demonstrates that our proposed method has better generalization ability among different network models. It is worth noting that the FLOPs and Params are far less than the teacher network. 

\vspace{-0.5cm}
\section{CONCLUSIONS}
\setlength{\baselineskip}{11.2pt}
\label{sec:typestyle}
In this paper, we propose LightVessel, an efficient framework tailored for contrary artery vessel segmentation based on knowledge distillation. We present FSD module is tailored to promote the semantic extraction ability of the student network. Meanwhile, the ASD module is associated with the pixel semantic space of ground truth to enhance fine-grained predictions of the student network. Our method is demonstrated in Clinic Contrary Artery Vessel Dataset, which achieves higher performance compared with various knowledge distillation methods.

\vfill\pagebreak



\end{document}